\documentclass[twocolumn,amsmath,floats,showpacs,nofootinbib]{revtex4}
\usepackage[usenames]{color}
\usepackage{graphicx}
\usepackage{epstopdf}
\usepackage{textcomp}
\usepackage{hyperref}
\usepackage{multirow}
\usepackage{tikz}
\usepackage{rotating}
\hypersetup{colorlinks=true}

\begin{document}

\title{Critical current density in the superconducting ceramic $\rm La_{1.8}Sr_{0.2}CuO_4$ in S-c-N point contacts}

\author{I.K. Yanson, L.F. Rybal'chenko, N.L. Bobrov, and V.V. Fisun}
\affiliation{B.Verkin Institute for Low Temperature Physics and Engineering, 47, Lenin Ave., 310164 Kharkov, Ukraine
Email address: bobrov@ilt.kharkov.ua}
\published {\href{http://fntr.ilt.kharkov.ua/fnt/pdf/13/13-8/f13-0873r.pdf}{Fiz. Nizk. Temp.}, \textbf{13}, 973 (1987)); (Sov. J. Low Temp. Phys., \textbf{13}, 501(1987)}
\date{\today}

\begin{abstract}Measurement of the current-voltage characteristics of point contacts makes it possible to study the properties of individual crystallites in a superconducting ceramic. The critical current density in the superconducting regions of the ceramic $\rm La_{1.8}Sr_{0.2}CuO_4$, with a size of the order of several tens of angstroms, is found to attain values of $10^8\ A/cm^2$, which are of the same order of magnitude as the pair-breaking current density, as evaluated from the formulas of the standard theory of superconductivity.


\end{abstract}

\maketitle
In superconducting oxides with high critical parameters ($T_c$, $H_{c2}$) the critical current density is usually many orders of magnitude lower than the pair-breaking current density, estimated from the formulas of the standard theory of conductivity \cite{1}. This is attributed either to the existence of weak contacts between the superconducting granules in such materials or by the very low vortex-pinning force, in the case of single crystals with a high degree of perfection \cite{2}.

Measurements of the I-V characteristics of point contacts whose size $d$ is of the order of several tens of angstr$\rm\ddot{o}$ms makes it possible to study the characteristics of a superconductor under conditions when a vortex structure is not formed and the effect of the nonuniformity of the order parameter at distances greater than $d$ can be disregarded because of the three-dimensional spreading of current from the region of the contact.

\begin{figure}[]
\includegraphics[width=8.5cm,angle=0]{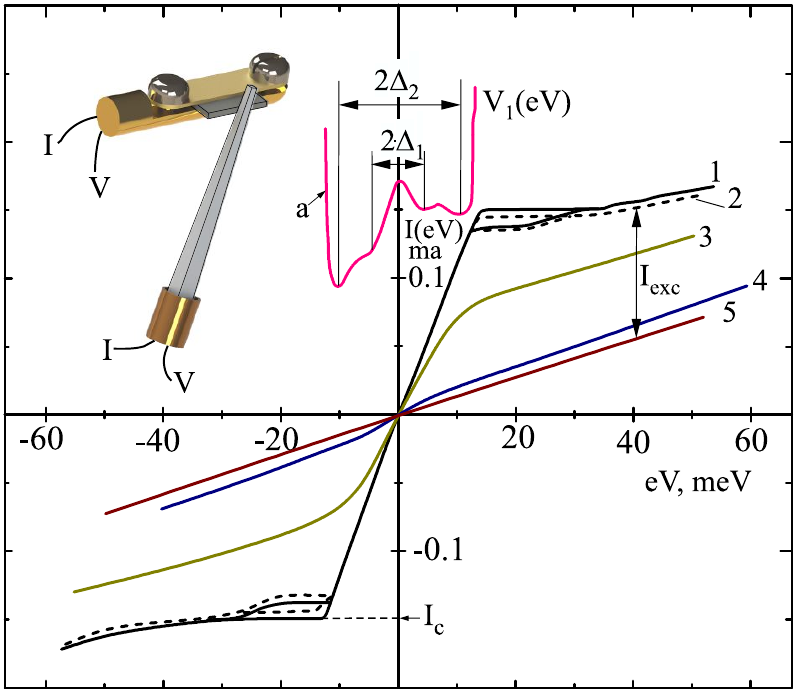}
\caption[]{Current-voltage characteristics of a copper-$\rm La_{1.8}Sr_{0.2}CuO_4$ ceramic point contact at 4.1~$K$ (1), 10.8~$K$(2), 19~$K$ (3), 24.2~$K$ (4), and 27.4~$K$ (5), the $dV/dI(V)$ characteristics at 10.8~$K$ (curve a), and the excess current $I_{exc}$ at $T = 4.1\ K$.}
\label{Fig1}
\end{figure}

Because of the low concentration of the superconducting phase in the $\rm La_{1.8}Sr_{0.2}CuO_4$ samples studied, it is very difficult to form an S-c-S point contact between two regions with high critical parameters. Using copper as the counterelectrode, we managed to prepare clamped $\rm Cu-La_{1.8}Sr_{0.2}CuO_4$ point contacts (see the inset to Fig.\ref{Fig1}), whose characteristics clearly displayed superconductivity at temperatures below $T_c^*\approx 27\ K$, close to the $T_c$ of the bulk sample.

The I-V curves shown in Fig.\ref{Fig1} were taken at different temperatures for one of the point contacts. For low biases and low temperatures the resistance $R_0$ is due to the copper electrode while for high $eV$ the I-V characteristic is similar to that in the normal state and the differential resistance $R_N$ is determined almost entirely by the resistance to contraction in the ceramic
($\rho_{ceram}\gg\rho_{Cu}$). The ratio $r=R_N/R_0$ varied over a wide range for different contacts and, apparently, is determined by the fraction of superconducting phase in the region near the contact on the side of the ceramic. The maximum values of $r$ (10-20) attained in the best contacts corresponded to a negligible contribution to $R_0$ from the resistance of the normal phase of the ceramic. At low temperatures the attainment of the critical current $I_c$ is accompanied by a jump in the voltage and hysteresis of the I-V characteristic. The superconductivity in the region of the contact is not destroyed completely, however, as is evidenced by the excess current $I_{exc}$, whose value is close to $I_c$. Apparently, a region of slippage of the order-parameter phase is formed and in this region normal quasiparticles
transporting current above $I_c$ are generated. The existence of "critical" and excess currents indicates that the superconductor near the contact is capable of carrying supercurrent with a density
$j_c\sim I_c/d^2\sim I_{exc}/d^2$.
\begin{figure}[]
\includegraphics[width=8.5cm,angle=0]{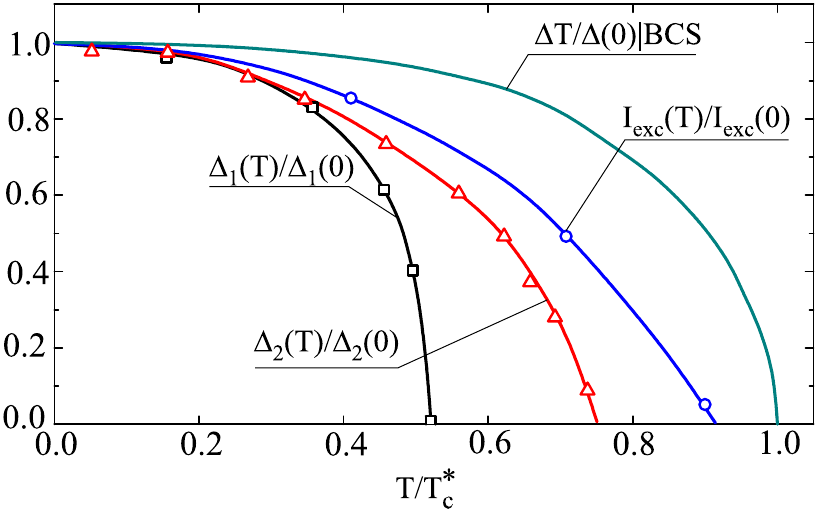}
\caption[]{Temperature dependences of $\Delta_1$, $\Delta_2$, and $I_{exc}$ in reduced coordinates.}
\label{Fig2}
\end{figure}

We evaluate the lower limit for the contact diameter, assuming that the electron transit in the copper edge is ballistic; $d\gtrsim (16\rho l/3\pi {{R}_{0}})^{1/2}$. For $R_0 = 83\ \Omega$ and
 $\rho l = 0.66\cdot 10^{-11}\ \Omega\cdot cm^2$ we obtain $d\gtrsim 37\ \text{\AA}$. On the other hand, assuming that the electron motion in the normal
ceramic is via diffusion ($l< d$), we obtain for the upper limit the value $d\lesssim \rho_{ceram}/2R_N\approx 120\ \text{\AA}$, where $R_N = 750\ \Omega$ and $\rho_{ceram}\approx 1.6\cdot 10^{-3}\ \Omega\cdot cm$ is the $\rm La_{1.8}Sr_{0.2}CuO_4$ bulk resistivity. Therefore, $1.0\cdot 10^8\ A/cm^2< j_c^{exp}< 1.1\cdot 10^9\ À/cm^2$. We compare this estimate with the pair-breaking current density $j_c =(\sqrt{2}/6\pi\sqrt{3})\times(cH_{cm}/\lambda)$. Substituting $H_{cm}\approx 4.5\cdot 10^3\ Oe$ and $\lambda\approx 2500\ \text{\AA}$ into this formula, we have $j_c^{theor}\approx 0.8\cdot10^8 \ A/cm^2$. An estimate of the pair-breaking current can also be obtained by using the experimental value of the gap $\Delta_2$, measured on the basis of the position of the $dV/dI$ minimum on the $eV$ axis (see curve a in Fig.\ref{Fig1}) and the carrier concentration $n\approx 10^{21}\ cm^{-3}$, giving $p_F\approx 3.2\cdot 10^{-20}\ g\cdot cm/sec$. For $\Delta_2(0)\approx 13\ meV$ we again obtain $j_c^{theor} =
en(\Delta/p_F)\approx 10^8\ A/cm^2$. As is known, the gap minima on the $dV/dI$ characteristics of point contacts with direct conduction are due to reflection of quasiparticles from the N-S boundary as a result of the mismatch of the electron parameters at the edges. From Fig.\ref{Fig1} (curve a) we see that $dV/dI$ has two minima, corresponding to two values of the gap. For the estimate we chose the large gap $\Delta_2$, since the transition to the segment of the I-V curve with the high resistance $R_N$ occurs near it. This transition means that the excess energy of the quasiparticles impinging on the superconductor exceeds the energy gap that determines the current-carrying capacity of the superconductor. We note that, according to our estimates, the contact diameter is comparable with the coherence length $\xi\approx 20\ \text{\AA}$. Vortices cannot come into being and move, therefore, in the vicinity of the contact.

From Fig.\ref{Fig2} we see that with increasing temperature the energy gaps $\Delta_1(T)$ and $\Delta_2(T)$ in reduced coordinates decrease more rapidly than $I_{exc}(T)$ does. These data support our assumption \cite{3} that the superconductivity of the ceramic is gapless in an appreciable temperature range below the temperature $T_c^*$, at which the nonlinearity of the I-V characteristic first appears near $V = 0$. Material with highly uniform superconducting properties is necessary in order to ascertain whether gapless superconductivity is an intrinsic property of ceramic superconductors or is due to the proximity effect in the presence of neighboring sections of normal phase.

We are grateful to B.I. Verkin for his attention and support in the performance of the work.


\begin{thebibliography}{}


\bibitem{1}V.V. Shmidt, Introduction to the Physic of Superconductors (in Russianl, Nauka, Moscow (1982).
\bibitem{2}B. Batlogg, J.P. Remeika, R.C. Dynes, et al.t "Structural instabilities and superconductivity in single crystal $\rm Ba(Pb,Bi)O_3$ in: Superconductivity in d- and f-band Metals, W. Buckel and W. Weber (eds.), Kernforschungszentrum Karlsruhe GmbH, Êarisruhef (1982), pp. 401-403.
\bibitem{3}I.K. Yanson, L.F. Rybal'chenko, V.V. Fisun, et al., \href{http://fntr.ilt.kharkov.ua/fnt/pdf/13/13-5/f13-0557r.pdf}{Fiz. Nizk. Temp.} \textbf{13}, 557 (1987) [Sov. J. Low Temp. Phys. \textbf{13}, 315 (1987)]; \href{https://arxiv.org/pdf/1701.01982.pdf}{arXiv:1701.01982}.

\end{thebibliography}
\end{document}